\documentclass[aps,prb,twocolumn,groupedaddress]{revtex4-1}
\usepackage[latin9]{inputenc}
\usepackage{graphicx}
\usepackage{epstopdf}

\makeatletter
 
 \@ifundefined{textcolor}{}
 {%
   \definecolor{BLACK}{gray}{0}
   \definecolor{WHITE}{gray}{1}
   \definecolor{RED}{rgb}{1,0,0}
   \definecolor{GREEN}{rgb}{0,1,0}
   \definecolor{BLUE}{rgb}{0,0,1}
   \definecolor{CYAN}{cmyk}{1,0,0,0}
   \definecolor{MAGENTA}{cmyk}{0,1,0,0}
   \definecolor{YELLOW}{cmyk}{0,0,1,0}
 }

\usepackage{dcolumn}
\usepackage{bm}

\makeatother

\begin{document}






\title{Theoretical 2D Raman band of strained graphene}

\author{Valentin N. Popov}

\affiliation{Faculty of Physics, University of Sofia, BG-1164 Sofia, Bulgaria}

\author{Philippe Lambin}

\affiliation{Research Center in Physics of Matter and Radiations, University of Namur (FUNDP), B-5000 Namur, Belgium}

\date{\today}
\begin{abstract}
We study the 2D Raman band of in-plane uniaxially strained graphene within a non-orthogonal tight-binding model. At non-zero strain, the obtained 2D band splits into two subbands at strain angles $0^{\circ}$ and $30^{\circ}$ or into three subbands at intermediate angles. The evolution of the 2D subbands is calculated systematically in the range of the accessible strains from $-1\%$ to $3\%$ and for the commonly used laser photon energy from $1.5$ eV to $3.0$ eV. The strain rate and dispersion rate of the 2D subbands are derived and tabulated. In particular, these two quantities show large variations up to $50\%$. The results on the 2D subbands can be used for detecting and monitoring strain in graphene for nanoelectronics applications. 
\end{abstract}
\maketitle


\section{Introduction}

The successful isolation of flakes of the two-dimensional crystal graphene and the measured high thermal and carrier mobility open new possibilities for application in nanoelectronics\cite{geim07} and nanophotonics.\cite{bona10} It is important to investigate the state of strain in graphene, because strain can effectively modify the electronic structure and phonon dispersion,\cite{mohr09} and thus change the transport properties of  this material .\cite{cast09} This can be done easily and in a non-destructive way by means of Raman spectroscopy.\cite{dres10}   

The Raman spectrum of perfect graphene is dominated by an intense first-order Raman band, the so-called G band, and a few second-order Raman bands.\cite{ferr06} The most intense one among the latter is observed at $\approx 2700$ cm$^{-1}$ and is termed the 2D band. This band is dispersive, i.e., it depends on the laser photon energy $E_{L}$. It is excited by double-resonant scattering processes,\cite{thom00,zoly11} favored by the specific cone-like electronic structure near the Fermi energy.  Raman measurements on strained graphene have been performed at different strain magnitude, strain direction, and laser photon energy.\cite{mohi09,huan10,fran11,yoon11} In all these studies, only one or two peaks of the 2D band could be resolved and their shift rates have been derived. In two similar measurements,\cite{huan10,yoon11} the determined strains differ by a factor of $2$, which shows that the calibration of strain can still be an issue. 

The dependence of the shift rate on strain can be obtained by calculation of the 2D band under various strain conditions. The complexity of these calculations at the \textit{ab-initio} level hinders the detailed investigation of the dependence of this band on strain and laser photon energy. The theoretical work has so far been limited to the calculation of the 2D peak position,\cite{fran11,yoon11,naru12} while the intensity of the peaks is either treated approximately,\cite{fran11,yoon11} or, in a limited number of cases, is calculated from first principles.\cite{naru12} Despite the success in modeling the 2D Raman band of graphene, a detailed investigation of its behavior under strain is mandatory for monitoring strain in this material by means of Raman spectroscopy.  

Recently, we have reported the predicted 2D band of graphene at strain magnitude $\epsilon=1\%$ and $E_{L}=2.5$ eV within a non-orthogonal tight-binding (NTB) model.\cite{popo12a} The very good agreement of our results with the available experimental data supports the applicability of this model to the 2D band of strained graphene.   

Here, we use the NTB model for large-scale calculations of the 2D band in graphene under uniaxial strain in the range of the accessible strain magnitudes and distinct strain directions, as well as for the common laser photon energies. The NTB model is presented in Sec. II. The obtained results are discussed in comparison with available experimental data in Sec. III. The paper ends up with conclusions (Sec. IV).  

\section{Computational details}

The NTB model\cite{popo04} uses Hamiltonian and overlap matrix elements obtained from an \textit{ab-initio} study.\cite{pore95} This model has no adjustable parameters, which is an advantage over the majority of tight-binding models. It also allows one to determine the total energy and forces on atoms. This feature is necessary for the relaxation of the atomic structure of graphene and for the calculation of the phonon dispersion. The latter is performed using a perturbative approach within the NTB model with electron-phonon matrix elements derived within the same model.\cite{popo10} The structure relaxation and phonon dispersion calculations require  the summation over the Brillouin zone of graphene. The phonon frequencies converge to within 1 cm$^{-1}$ on increasing the size of the Monkhorst-Pack mesh of points up to $40\times40$. The calculated high-frequency phonon branches are overestimated by the NTB model. \cite{popo10} However, very good agreement with experiment can be reached by scaling the phonon frequencies of these branches by a factor of $0.9$. \cite{mala09a} For this reason, \textit{the phonon frequencies are systematically scaled by a factor of $0.9$ for all calculations of the 2D band in the paper.}

The Raman intensity of the 2D band is calculated using fourth-order terms in quantum-mechanical perturbation theory.\cite{popo12} The required electron-photon and electron-phonon matrix elements, and electronic linewidth are obtained within the NTB model. The expression for the intensity contains summation over all electronic states and over all phonons in the entire Brillouin zone. The convergence of the Raman shift of the 2D band within 1 cm$^{-1}$ is achieved  by increasing the size of the Monkhorst-Pack mesh of points up to $800\times800$. 

\begin{figure}[t]
\includegraphics[width=60mm]{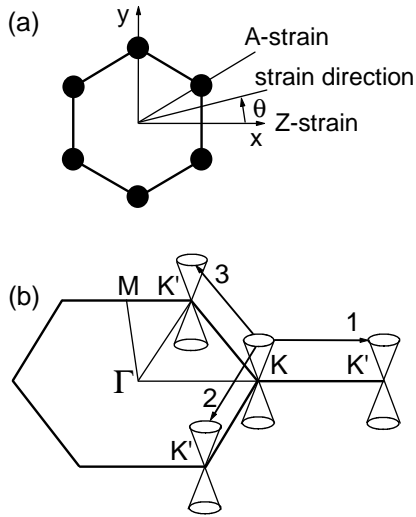} 
\caption{(a) The hexagonal atomic structure of graphene. The atoms are depicted by solid circles and the solid lines are the interatomic bonds. The strain direction is defined by the angle $\theta$ relative to the $x$ axis, which is chosen along a zigzag of carbon bonds. The limiting cases of Z-strain and A-strain are also shown. (b) Schematic representation of a double-resonant Raman scattering process, which contributes to the 2D band. An electron is scattered between the cone-like conduction bands at the K and K$^{'}$ points of the hexagonal Brillouin zone of graphene by a K point phonon with a wavevector, denoted by an arrow. The electron is scattered back by a phonon with an opposite wavevector. The three scattering paths KK$^{'}$ are denoted by 1, 2, and 3. Similarly, holes can be scattered by phonons between cone-like valence bands at the K and K$^{'}$ points. }
\end{figure}

\section{Results and Discussion}

We calculate the electronic structure and phonon dispersion of the 2D band of uniaxially strained graphene at strain magnitude $\epsilon =$ $-1\%$, $1\%$, $2\%$, and $3\%$ and strain direction, defined by the angle $\theta$ relative to a zigzag of carbon bonds, with values $\theta =$ $0^{\circ}$, $10^{\circ}$, $20^{\circ}$, and $30^{\circ}$ (Fig.~1a). The interval from $\theta =$ $ 0^{\circ}$ (Z-strain) to $\theta =$ $30^{\circ}$ (A-strain) covers all distinct strain directions in graphene. The electronic and phonon dispersions of strained graphene have been discussed a lot in the literature (see, e.g., Ref.~\onlinecite{mohr09}). Previously, we have demonstrated that both dispersions are correctly described by the NTB model\cite{popo12a} and for this reason, here, we focus on the study of the 2D band and its dependence on strain. We consider laser photon energies $E_{L} = $$1.5$, $2.0$, $2.5$, and $3.0$ eV, and restrict ourselves to parallel and perpendicular light polarizations relative to the strain direction for both polarizer and analyzer. Finally, we are concerned only with the Stokes part of the Raman spectra.
 
In order to facilitate the understanding of the following results, we describe briefly the double-resonant processes. In such a process, an electron (hole) is scattered between states of the cone-like conduction (valence) bands at the K and K$^{'}$ points of the Brillouin zone of graphene  (so-called Dirac cones)  by two phonons with opposite wavevectors close to the K point  (Fig.~1b). The resulting 2D band is centered at roughly twice the frequency of the phonons. Each K point has three neighboring K$^{'}$ points and therefore there are three scattering paths KK$^{'}$. Due to the hexagonal symmetry of perfect graphene, the scattering along the three paths yields equal contribution to the Raman shift and intensity of the 2D band.

It has been debated in the literature about the relative contribution to the 2D band from phonons close to the $\Gamma$K and KM directions in the Brillouin zone.\cite{naru12} Such phonons and the associated processes are often called ``inner'' and ``outer", respectively. Some authors have concluded that the contributions from these processes are comparable,\cite{naru11,fran11} while others have found that the contribution from ``outer'' processes is negligible compared to that from ``inner'' ones.\cite{ vene11,yoon11} It has also been argued that the phonon wavevector for a seemingly ``outer'' process can be transformed by a reciprocal lattice vector to a phonon wavevector for an equivalent ``inner'' process, ``thereby dismissing the notion of inner and outer processes".\cite{naru12} In a recent study, we have proposed a quantitative criterion for distinction between the two processes. \cite{popo12} We have noted that (1) the phonons with largest contribution to the 2D band belong exclusively to the TO phonon branch and (2) the TO branch is the fifth one close to the $\Gamma$K direction and the sixth one close to the KM direction (in order of increasing frequency). Based on these observations, we have defined unambiguously ``inner'' and ``outer'' processes as processes due to phonons from the fifth and sixth phonon branches close to the K point, respectively. Using this definition, we have shown that in unstrained graphene the ratio of the contributions to the 2D band from ``inner'' and ``outer'' processes is about 10.

The calculated 2D band of unstrained graphene has a symmetric shape (Fig.~2, dotted line). This result is a direct consequence of the dominant contribution from ``inner'' processes and is in agreement with the observed symmetric shape of the 2D band.\cite{huan10}

\subsection{Strain dependence of the Raman intensity of the 2D subbands}

\begin{figure}[t]
\includegraphics[width=85mm]{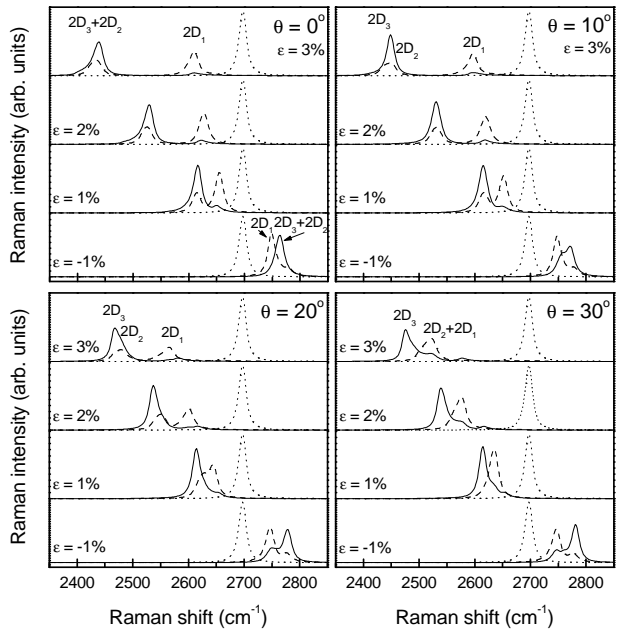} 
\caption{Calculated 2D band at $\theta = 0^{\circ}$ (Z-strain), $10^{\circ}$, $20^{\circ}$, and $30^{\circ}$ (A-strain), at $E_{L} = 2.0$ eV, and for parallel/perpendicular light polarization (solid/dashed lines) in comparison with that at zero strain (dotted line). In most cases, two Raman features are clearly seen. They correspond to the experimentally observed 2D$^{-}$ and 2D$^{+}$ bands.}
\end{figure}

The calculated 2D Raman spectra at different strain magnitude and at $E_{L}$ = $2.0$ eV, for parallel and perpendicular light polarization are shown in Fig.~2. The spectra at $E_{L}$ = $1.5$, $2.5$, and $3.0$ eV are also calculated but are not provided here.  In all cases, two or three 2D subbands are present. In order to explain their origin, we note that in strained graphene, the electronic and phonon dispersions are modified differently along the three paths KK$^{'}$. Therefore,  the 2D band splits into two or three subbands depending on the strain direction. The subbands, coming from paths 1, 2, and 3, are denoted here by 2D$_{1}$, 2D$_{2}$, and 2D$_{3}$, respectively. These subbands are mainly due to ``inner'' processes, while the contributions from ``outer'' processes give rise to minor peaks and kinks, which deform slightly the shape of the 2D subbands. 

The calculated spectra for Z-strain graphene are given in Fig.~2 (upper left panel). In this case, scattering paths 2 and 3 are equivalent (see Fig.~1b). Under tensile strain ($\epsilon>0$) the 2D subbands are red-shifted, while under compressive strain ($\epsilon<0$) they are blue-shifted. The red (blue) shift increases with increasing strain (in absolute value). There are only two distinct 2D subbands: the composite 2D$_{3}$+2D$_{2}$ band and the 2D$_{1}$ subband. The splitting between them increases with increasing $\epsilon$. For parallel polarization, only the composite band is clearly seen, while, for perpendicular polarization, both subbands can be observed in the Raman spectrum. For tensile strain, the composite band has smaller Raman shift than the 2D$_{1}$ subband and vice versa for compressive strain.

For the intermediate values of the strain direction angle $\theta =$ $10^{\circ}$ and $20^{\circ}$, there are three 2D subbands, as seen in Fig.~2 (upper right and lower left panels). The separation between the 2D$_{3}$ and 2D$_{1}$ subbands increases with increasing $\epsilon$ and decreases with increasing $\theta$. At $\theta =$ $10^{\circ}$ and $\epsilon > 0$, and $\theta =$ $20^{\circ}$ and $\epsilon < 0$, the 2D$_{3}$ and 2D$_{2}$ subbands overlap considerably and under such strains and angles only two 2D peaks can be observed: one peak, corresponding to the overlapping subbands 2D$_{3}$ and 2D$_{2}$, and another peak, corresponding to the subband 2D$_{1}$. Figure~2 (lower right panel) shows that for $\theta =$ $30^{\circ}$ only two Raman features are present, namely, the 2D$_{3}$ subband and the composite 2D$_{2}$+2D$_{1}$ band. The red (blue) shift of the subbands changes with $\epsilon$ similarly to the cases of  the values of $\theta$ considered above. 
 
Next, we compare the shape of the obtained 2D band to available experimental Raman spectra of strained graphene. As a general feature, two 2D Raman peaks are usually observed. They are termed 2D$^{-}$ and 2D$^{+}$ band, the latter having larger shift than the former.\cite{huan10,fran11,yoon11} Measurements on Z- and A-strain graphene at tensile strains up to $3.2\%$ and $3.8\%$, respectively, have been performed at $E_{L}=2.33$ eV (Ref.~\onlinecite{huan10}). The measured Raman spectra show mainly one intense peak and a minor peak or a kink, especially, for Z-strain at large $\epsilon$.  Such a behavior is qualitatively reproduced by our spectra in Fig.~2 (upper left and lower right panels). The asymmetry of the 2D band shape, namely, higher minor peak for perpendicular polarization than for parallel polarization, is also predicted here. In another experiment,\cite{fran11} Raman measurement have been performed on a graphene flake at $\theta = 18.4^{\circ}$ and tensile strain $\epsilon$ up to $\approx0.62$ ($0.58$)$\%$, and on another flake at $\theta = 10.8^{\circ}$ and $\epsilon$ up to $\approx 1.17$ ($1.17$)$\%$ for parallel (perpendicular) light polarization; in both cases $E_{L} = 1.58$ eV. In neither case, the 2D band could be fitted with more than two Lorentzian functions, because of the small 2D$_{2}-$2D$_{3}$ band splitting at such relatively small $E_{L}$. Our spectra in Fig.~2  are in accord with these observations. In Ref. \onlinecite{fran11}, the overall symmetric 2D band of unstrained graphene has been fitted with two Lorentzian functions, attributed to ``inner'' and ``outer'' processes. It has been concluded that the two types of processes give comparable contributions to the band. In a recent paper,\cite{yoon11} close to Z- and A-strain graphene samples have been studied by Raman spectroscopy at strains up to $1.19$ ($1.15$)$\%$, at $E_{L} = 2.41$ eV and for parallel (perpendicular) light polarization. By analyzing the 2D band shape and splitting, it has been argued that this band comes mainly from ``inner'' processes. Our simulated spectra exhibit similar 2D band splitting and shape, as the latter experimental ones, in support of the relatively small contribution from ``outer'' processes. 

\begin{table*}[t]
\caption{\label{tab:table1}
Calculated shift rates of the subbands 2D$_{1}$ and 2D$_{3}$  (in cm$^{-1}$/$\%$), corresponding to the experimentally observed subbands 2D$^{+}$ and 2D$^{-}$, in comparison with available experimental and \textit{ab-initio} data. The relative deviation of the NTB values from the experimental (\textit{ab-initio}) ones is given in the columns labeled ``Rel. dev." The second row for each couple of values for E$_{L}$ and $\theta$ contains the NTB values scaled by $0.8$ and the corresponding relative deviation (see text).}
\begin{ruledtabular}
\begin{tabular}{ccllllll}
$E_{L}$&$\theta$&\multicolumn{3}{c}{2D$_{1}$ (2D$^{+}$)}&\multicolumn{3}{c}{2D$_{3}$ (2D$^{-}$)}\\
eV&&Expt. (\textit{ab-initio})&NTB&Rel. dev.&Expt. (\textit{ab-initio})&NTB&Rel. dev.\\
\hline
$2.33$&$0^{\circ}$&$-16.3$\footnotemark[1]&$-32.4$&$99\%$&$-29.7$\footnotemark[1]&$-81.1$&$173\%$\\
&&&$-25.9$&$59\%$&&$-64.9$&$119\%$\\
$2.33$&$30^{\circ}$&$-21.7$\footnotemark[1]&$-54.3$&$150\%$&$-30.5$\footnotemark[1]&$-80.3$&$163\%$\\
&&&$-43.4$&$100\%$&&$-64.2$&$110\%$\\
$1.58$&$18.4^{\circ}$&$\approx-30$ ($-49.4$)\footnotemark[2] &$-46.8$&$56\%$ ($$-5\%)&$-46.8$ ($-57.4$)\footnotemark[2] &$-77.4$&$65\%$ ($35\%$)\\
&&&$-37.4$&$25\%$ ($$-24\%)&&$-61.9$&$35\%$ ($8\%$)\\

$2.41$&$\approx0^{\circ}$&$-26.0$ ($-24$)\footnotemark[3] &$-31.8$&$22\%$ ($31\%$)&$-67.8$ ($-66$)\footnotemark[3]&$-81.0$&$19\%$ ($23\%$)\\
&&&$-25.4$&$-2\%$ ($6\%$)&&$-64.8$&$-4\%$ ($-2\%$)\\
$2.41$&$\approx30^{\circ}$&$-44.1$ ($-43$)\footnotemark[3] &$-53.5$&$21\%$ ($24\%$)&$-63.1$ ($-70$)\footnotemark[3] &$-81.1$&$29\%$ ($16\%$)\\
&&&$-42.8$&$-3\%$ ($0\%$)&&$-64.9$&$3\%$ ($-7\%$)\\
\end{tabular}
\end{ruledtabular}
\footnotetext[1]{Experimental data from Ref.~\onlinecite{huan10}.}
\footnotetext[2]{Experimental (\textit{ab-initio}) data from Ref.~\onlinecite{fran11}.}
\footnotetext[3]{Experimental (\textit{ab-initio}) data from Ref.~\onlinecite{yoon11}.}

\end{table*}

The 2D Raman bands can be characterized by their linewidth and Raman shift. The calculated full width at half maximum increases steeply with $E_{L}$ (in eV)  at $\epsilon=0$ as $13.2+0.067E_{L}^{5}$ (in cm$^{-1}$). The predicted linewidth of $17.8$ cm$^{-1}$ is more than twice smaller than the measured one of $\approx40$ cm$^{-1}$  at $E_{L}=2.33$ eV  (Ref.~\onlinecite{huan10}). Our linewidth of  $15$ cm$^{-1}$ is by $50\%$ smaller than the averaged one of $24$ cm$^{-1}$ for two samples with $\theta \approx20\%$, measured at $E_{L}=1.58$ eV (Ref.~\onlinecite{fran11}). At $E_{L}=2.41$ eV, we calculate the value of $18.6$ cm$^{-1}$, which is about $20\%$ smaller than the experimental one of $\approx22$ cm$^{-1}$  (Ref.~\onlinecite{yoon11}). The disagreement of the theoretical linewidth with the observed one can be due to electron scattering mechanisms, unaccounted for here, or to any extrinsic mechanism that reduces the lifetime of the phonon.

The theoretical linewidth increases weakly with $\epsilon$ at the rate of a few cm$^{-1}$/$\%$ at $E_{L}=2.0$ eV (see Fig. 2). This behavior corresponds to that reported in Ref.~\onlinecite{huan10,yoon11}. In Ref.~\onlinecite{fran11}, the measured 2D band at tensile strain has been fitted by a single Lorentzian and the obtained linewidth has been found to increase steeply with $\epsilon$. In this case, the linewidth is roughly equal to the sum of the halfwidths of the lower and upper 2D subbands, and the separation between them. Our estimated separation of $11$ cm$^{-1}$ corresponds well to the increase of the measured linewidth of $\approx 12$ cm$^{-1}$ at $\epsilon \approx 0.5\%$.

\subsection{Dependence of the Raman shift of the 2D subbands on $\epsilon$}

First, we compare the calculated red shift of the 2D band to the available experimental data. One of the early measurement of these shifts has been performed on graphene flakes in polymer films using two- and four-point bending setups at $E_{L}=2.41$ eV (Ref.~\onlinecite{mohi09}). No splitting of the 2D band has been observed and the red shift per unit strain (so-called \textit{shift rate}) of $-64$ cm$^{-1}$/$\%$ has been obtained. A three-point setup for Raman measurements on polymer-imbedded graphene flakes has been used for systematic investigation of A- and Z-strain samples at $E_{L}=2.33$ eV.\cite{huan10} The shift rates of the two  observed peaks of the 2D band are $-16.3$ and $-29.7$ cm$^{-1}$/$\%$ for Z-strain and $-21.7$ and $-30.5$ cm$^{-1}$/$\%$ for A-strain. These values are at least twice smaller than those of Ref.~\onlinecite{mohi09}. Later on, the analysis of the Raman data of strained graphene at $\theta=18.4^{\circ}$, measured at $E_{L}=1.58$ eV, has allowed to derive the value of $\approx -30$ cm$^{-1}$/$\%$  (averaged)   ($-46.8$ cm$^{-1}$/$\%$) for the shift rate of the higher (lower) 2D component.\cite{fran11} The shift rates of $-49.4$ and $-57.4$ cm$^{-1}$/$\%$, calculated from first principles, overestimate the experimental values by more than $20\%$ (in absolute value). In another paper, the pairs of shift rates $-44.1$ and $-63.1$ cm$^{-1}$/$\%$ (near-A-strain) and $-26.0$ and $-67.8$ cm$^{-1}$/$\%$ (near-Z-strain), measured at $E_{L}=2.41$ eV, have been reported.\cite{yoon11} The \textit{ab-initio} calculations for the two samples have yielded for the pairs of shift rates $-43$ and $-70$ cm$^{-1}$/$\%$, and $-24$ and $-66$ cm$^{-1}$/$\%$, respectively. These values reproduce well the shift rates of the 2D subbands, except for the lower subband at A-strain, for which the predicted rate is larger (in absolute value) than the measured one by about $10\%$.

Table I shows that the NTB shift rates are systematically larger than the available experimental data, the overestimation ranging from $19\%$ to $173\%$. Similarly to the case of the G band of strained graphene,\cite{popo12} the larger shift rates of the 2D subbands  can be reduced and agreement with experiment can be reached by scaling them by a single factor. For the determination of the scaling factor it is reasonable to choose experimental data with observed 2D band splitting and available first-principles predictions. Moreover, the reliability of the scaling factor will depend on the agreement between the experimental and theoretical data. Following these arguments, we use the experimental shift rates of Ref.~\onlinecite{yoon11} because they agree well with the \textit{ab-initio} ones. From the average overestimation of the data from the latter paper, we derive a scaling factor of $\approx0.8$. The value of this factor can be obtained more accurately by using more precise experimental and theoretical data.  In this paper, we accept this scaling factor, and \textit{the red (blue) shift will be scaled by $0.8$} everywhere below, in addition to the scaling of the phonon frequencies by $0.9$. For brevity, the Raman shift with scaled red (blue) shift will be referred to as the \textit{scaled Raman shift}. As evidenced from Table~I, our scaled shift rates overestimate the experimental data of Ref.~\onlinecite{huan10} and Ref.~\onlinecite{fran11} in average by $\approx100\%$  and $\approx30\%$, respectively. 

\begin{figure}[t]
\includegraphics[width=80mm]{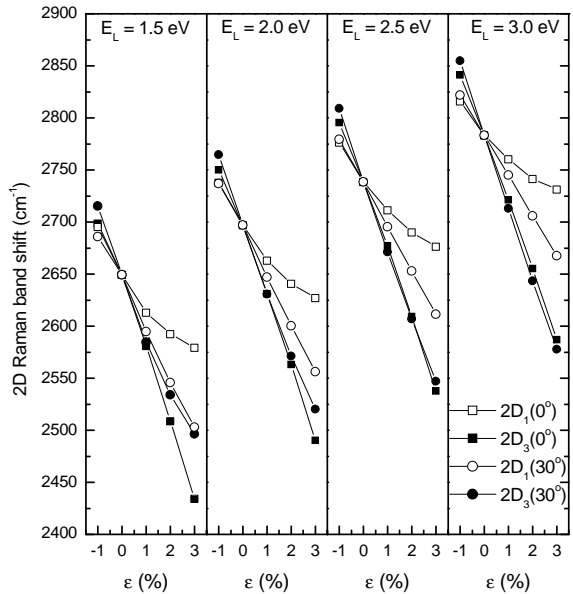} 
\caption{Scaled Raman shift of subbands 2D$_{1}$ and 2D$_{3}$ (symbols) as a function of the strain $\epsilon$ at $\theta=0^{\circ}$ and $30^{\circ}$, and at different values of $E_{L}$.}
\end{figure}

The scaled Raman shifts of subbands 2D$_{1}$  and 2D$_{3}$, derived from our spectra, are presented in Fig.~3 as a function of $\epsilon$ at different values of $E_{L}$ and $\theta$. In all cases, except for the 2D$_{1}$ subband at $\theta=0^{\circ}$, the shifts are quasi-linear in $\epsilon$. The shift rate is approximated here by the slope of the linear curve fitted to the data points in Fig.~3. We note that, in all cases, a linear curve is sufficient to fit the data except for the 2D$_{1}$ subband at $\theta=0^{\circ}$, where a polynomial of at least second degree is needed for a satisfactory fit. The shift rates, derived from Fig.~3, are presented in Table~II. The shift rate for intermediate values of $E_{L}$ and $\theta$ can be obtained by interpolation.

Finally, we calculate the Raman spectra of Z- and A-strain graphene at $E_{L}=2.33$ eV and light polarization as for the measured spectra in Fig.~3 of Ref.~\onlinecite{huan10}. Namely, the incident laser light is polarized parallelly or perpendicularly to the strain direction and the scattered light is observed at an angle of $0^{\circ}$, $45^{\circ}$, or $90^{\circ}$ relative to the incident light polarization. Since our strain rates overestimate the experimental ones of Ref.~\onlinecite{huan10} by $\approx100\%$, our calculations are performed at twice smaller strain magnitudes than the reported ones, i.e., at $\epsilon=1.9\%$ (Z-strain) and $\epsilon=1.3\%$ (A-strain). The obtained spectra (Fig.~4) agree qualitatively with the experimental ones. Major discrepancy is observed at Z-strain, where the predicted 2D$_{1}$-2D$_{3}$ splitting clearly resolved in perpendicular geometry is $\approx50\%$ larger than the experimental one. Obviously, more experimental and theoretical efforts are necessary for the unambiguous determination of the shift rates of the 2D band.

\begin{table}[t]
\caption{\label{tab:table2}
Shift rates (in cm$^{-1}$/$\%$) of the subbands 2D$_{1}$ and 2D$_{3}$ at different values of $E_{L}$ (in eV) and $\theta$, derived from Fig.~3 }
\begin{ruledtabular}
\begin{tabular}{cllllllll}
 $E_{L}$&\multicolumn{2}{c}{$\theta=0^{\circ}$}&%
\multicolumn{2}{c}{$\theta=10^{\circ}$}&%
\multicolumn{2}{c}{$\theta=20^{\circ}$}&\multicolumn{2}{c}{$\theta=30^{\circ}$}\\
&2D$_{1}$&2D$_{3}$&2D$_{1}$&2D$_{3}$&2D$_{1}$&2D$_{3}$&2D$_{1}$&2D$_{3}$\\
\hline
$1.5$&$-25.3$&$-66.9$&$-32.0$&$-64.9$&$-38.7$&$-61.0$&$-47.0$&$-55.4$\\
$2.0$&$-27.8$&$-65.4$&$-31.2$&$-64.7$&$-37.6$&$-62.9$&$-45.8$&$-61.4$\\
$2.5$&$-24.9$&$-64.7$&$-27.2$&$-65.1$&$-33.4$&$-65.6$&$-42.2$&$-65.7$\\
$3.0$&$-21.1$&$-63.7$&$-23.4$&$-67.5$&$-29.9$&$-68.8$&$-38.6$&$-69.4$\\
\end{tabular}
\end{ruledtabular}

\end{table}

\begin{figure}[t]
\includegraphics[width=85 mm]{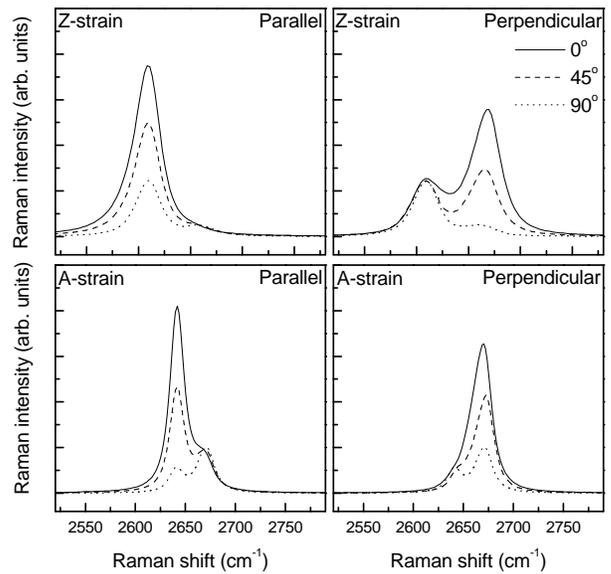} 
\caption{Calculated Raman spectra at Z-strain ($\epsilon=1.9\%$) and A-strain ($\epsilon=1.3\%$), at $E_{L}=2.33$ eV and parallel or perpendicular incident light polarization. The scattered light is polarized at an angle of $0^{\circ}$, $45^{\circ}$, or $90^{\circ}$ relative to the incident light polarization. The obtained spectra are in qualitative agreement with those reported in Fig.~3 of Ref.~\onlinecite{huan10} }
\end{figure}

\subsection{Dependence of the Raman shift of the 2D subbands on $E_{L}$}

The scaled Raman shifts of subbands 2D$_{1}$ and 2D$_{3}$, derived from our spectra, are shown in Fig.~5 as a function of $E_{L}$ at different $\epsilon$ and $\theta$. In all cases, except for the 2D$_{3}$ subband at $\theta = 30^{\circ}$, this function is quasi-linear and can be characterized by its first derivative, i.e., the slope of the corresponding curve. This quantity describes the dispersion of the 2D band and is sometimes referred to as the \textit{dispersion rate}. It is approximated here by the slope of the linear curve fitted to the data. We note that in the case of subband 2D$_{3}$ at $\theta = 30^{\circ}$, a second-degree polynomial fits the data more accurately. 

Table~III presents the dispersion rate of the 2D subbands, derived from Fig.~5, in comparison with that for unstrained graphene of $89$ cm$^{-1}$/eV (Ref.~\onlinecite{popo12a}). In the interval of the studied tensile strains, the dispersion rate of the 2D$_{1}$ subband increases up to $20\%$. The dispersion rate for the 2D$_{3}$ subband at $\theta=0^{\circ}$ and $10^{\circ}$ increases slightly, whereas it shows a decrease down to $20\%$ at $\theta=20^{\circ}$ and by almost $50\%$ at $\theta=30^{\circ}$. Therefore, the variation of the Raman shift with $E_{L}$ is not negligible and has to be taken into account for the correct assignment of the 2D components of strained graphene.

The dependence of the dispersion rate of the 2D band of strained graphene on $E_{L}$ has been investigated experimentally in the case of Z-strain.\cite{huan10} The reported rates are $93$, $99$, and $106$ cm$^{-1}$/eV for the 2D band of unstrained graphene and the 2D$_{3}$ and 2D$_{1}$ subbands of strained graphene at $\epsilon=2.6\%$, respectively. It has been argued that this strain might be about three times larger than the actual one.\cite{naru12} As shown above, our strain rates overestimate the experimental ones of Ref.~\onlinecite{huan10} by $\approx100\%$. Therefore, we compare the measured rates to the calculated ones of $95$ and $99$ cm$^{-1}$/eV for strained graphene at $\epsilon=1.3\%$. Our rates for unstrained and strained graphene underestimate the experimental ones roughly by $5\%$. The clarification of the origin of this discrepancy requires further theoretical and experimental efforts.

\begin{figure}[t]
\includegraphics[width=80mm]{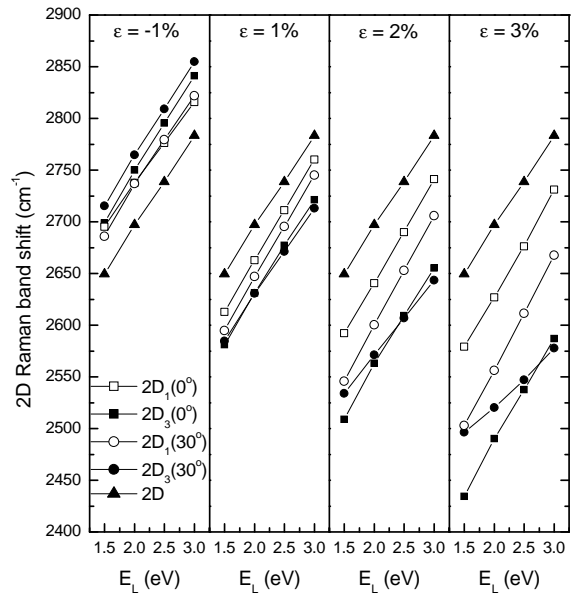} 
\caption{Scaled Raman shift of subbands 2D$_{1}$ and 2D$_{3}$ as a function of $E_L$ at different $\epsilon$ and at $\theta=0^{\circ}$ and $30^{\circ}$. The triangles are the shifts of the 2D band of unstrained graphene. The lines are guides to the eye.}
\end{figure}

\begin{table}[t]
\caption{\label{tab:table3}
Dispersion rate (in cm$^{-1}$/eV) of subbands 2D$_{1}$ and 2D$_{3}$ at different $\epsilon$ (in $\%$) and $\theta$, derived from Fig.~5.}
\begin{ruledtabular}
\begin{tabular}{rllllllll}
 $\epsilon$&\multicolumn{2}{c}{$\theta=0^{\circ}$}&%
\multicolumn{2}{c}{$\theta=10^{\circ}$}&%
\multicolumn{2}{c}{$\theta=20^{\circ}$}&\multicolumn{2}{c}{$\theta=30^{\circ}$}\\
&2D$_{1}$&2D$_{3}$&2D$_{1}$&2D$_{3}$&2D$_{1}$&2D$_{3}$&2D$_{1}$&2D$_{3}$\\
\hline
$-1$& $79$&$95$&     $81$&$85$&  $87$&$93$&$90$& $93$\\
$0$ &$89$& $89$&     $89$&$89$& $89$& $89$&$89$& $89$\\
$1$ &$99$&$93$&    $100$&$91$&$101$&$86$&$101$&$85$\\
$2$ &$100$&$98$&  $104$&$91$&$107$&$79$&$108$&$71$\\
$3$ &$102$&$102$&$105$&$91$&$110$&$70$&$112$&$50$\\
\end{tabular}
\end{ruledtabular}
\end{table}

Within the simple model of linear electronic and phonon dispersions in the vicinity of the K point, the dispersion rate has been expressed as $(2/\hbar)$$(v_{ph}/v_{F})$, where $v_{ph}$ and $v_{F}$ are the  phonon group and electronic velocities near the K point, respectively, and $\hbar$ is Planck's constant over $2\pi$.\cite{huan10} In the case of tensile Z-strain, it has been argued that $v_{F}$ decreases with increasing strain, the decrease being larger along path $1$ than along paths $2$ and $3$. Based on these arguments and the assumption for isotropic phonon group velocity, it has been concluded that the expression for the dispersion rate above describes the observed behavior of this quantity.\cite{huan10} On the other hand, in Ref.~\onlinecite{mohr10,yoon11}, it has been demonstrated that $v_{ph}$ of strained graphene is anisotropic. The NTB calculations on Z-strained graphene show that, with increasing tensile strain, $v_{F}$ remains almost unchanged and isotropic (the Dirac cones are displaced almost undeformed), while $v_{ph}$ increases and becomes anisotropic. Moreover, the latter increases more for path $1$ than for paths $2$ and $3$. The NTB velocities, substituted in the expression for the dispersion rate, yield within a few percent the theoretical one in Table III. The fact that contradicting theoretical arguments predict the same experimental observation indicates that more theoretical work is needed to reveal the correct dependence of the Fermi and phonon group velocities on strain.

\subsection{Dependence of the Raman shift of the 2D subbands on $\theta$}

\begin{figure}[t]
\includegraphics[width=85 mm]{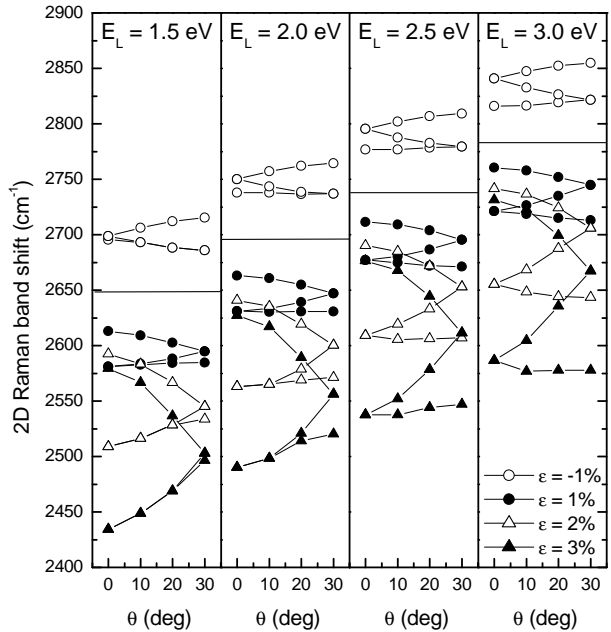} 
\caption{Scaled Raman shift of the 2D subbands (symbols)  as a function of $\theta$ at different values of $\epsilon$ and $E_{L}$. The three branches at a given strain magnitude are the shifts of subbands 2D$_{1}$, 2D$_{2}$, and 2D$_{3}$ in order of increasing (decreasing) shift at $\epsilon<0$ ($\epsilon>0$). The horizontal lines are drawn at the Raman shift of unstrained graphene. }
\end{figure}

Figure~6 presents the Raman shift of the 2D Raman subband,  extracted from the calculated spectra. For all sets of values of $\epsilon$, $\theta$, and $E_{L}$, the shift of the 2D$_{2}$ subband is between that of the 2D$_{1}$ and 2D$_{3}$ subbands. At $\theta=0^\circ$ the shifts of the 2D$_{2}$ and 2D$_{3}$ subbands are equal and these subbands contribute to the composite band 2D$_{2}$ $+$ 2D$_{3}$. Similarly, at $\theta=30^\circ$ the shifts of the 2D$_{1}$ and 2D$_{2}$ subbands are equal and they should be observed as a single composite band 2D$_{1}$ $+$ 2D$_{2}$. At small $E_{L}$, the splitting of subbands 2D$_{2}$ and 2D$_{3}$ is small enough to make them difficult to separate in Raman experiments. We also notice that the curves of points for $\epsilon = -1\%$ and $\epsilon = 1\%$ are almost centrosymmetric. This can be explained by the fact that a small tensile Z-strain deforms the atomic structure of graphene similarly to a small compressive A-strain and vice versa.

The derived shifts form specific patterns, which can be fitted with a polynomial in $\sin(\theta)$ and  $\cos(\theta)$. The origin of such a dependence can be found in the fact that, upon a small uniaxial deformation, any vector {\bf x} transforms into the vector $(1+\bm \epsilon')\bf x$, where $\bm \epsilon'$ is the strain tensor at strain angle $\theta$. This tensor can be obtained from the strain tensor $\bm \epsilon$ at $\theta=0^\circ$ by a rotation $\theta$ by means of an orthogonal transformation, defined by an orthogonal tensor $\bm R$: $\bm \epsilon^{'}=\bm R^{-1}\bm \epsilon \bm R$.\cite{mohr10} In the chosen coordinate system, $\bm \epsilon$ is represented by a $2\times2$ matrix with nonzero diagonal elements $\epsilon_{xx}=\epsilon$ and $\epsilon_{yy}=-\nu\epsilon$ ($\nu$ is the Poisson's ratio). The tensor $\bm R$ is represented by a $2\times2$ matrix with elements $\pm\sin(\theta)$ and  $\cos(\theta)$. The dynamical matrix for the strained structure can be expanded in a series in $\epsilon$, which consists of terms of even degree in $\sin(\theta)$ and $\cos(\theta )$. Therefore, the phonon frequency can be written as a polynomial of these trigonometric functions of even degree. In Ref. \onlinecite{popo12a}, a single function $\sin^4(\theta + \theta_{0})$ ($\theta_{0}$ is a proper constant phase shift, depending on the subband) was found sufficient to achieve a fair fit to the Raman shifts of the 2D subbands at $\epsilon =1\%$. For the considered sets of values of $\epsilon$ and $E_{L}$, a single trigonometric function gives an unsatisfactory fit to the 2D subbands, while a fair fit requires a polynomial of $\sin^2(\theta + \theta_{0})$ of at least fourth degree. Here, we did not attempt to fit such a polynomial to our data because the data points are not sufficient for the accurate determination of the polynomial coefficients. For the practical use of our results, an interpolation between the calculated data points should be done.

\section{Conclusions}

We have presented a detailed study of the 2D Raman band of strained graphene within a non-orthogonal tight-binding model. We have obtained the dependence of the Raman shift of the 2D subbands on the strain magnitude and direction, and on the laser photon energy. The derived red (blue) shift of the subbands systematically overestimates the available experimental data, which can be cured by using a constant scaling factor of $0.8$. The shift rate and the dispersion rate of the 2D band change by up to $50\%$ with varying the strain magnitude and direction, and laser photon energy. This result underlines the importance of the theoretical and experimental investigation of the behavior of the 2D Raman band of strained graphene under different experimental conditions. Our results for the scaled Raman shift, shift rate, and dispersion rate of the 2D band can be used together with the predictions for the shift rate of the G band\cite{popo12a} for monitoring strain in graphene by means of Raman spectroscopy. 

\acknowledgments

V.N.P. acknowledges financial support from the University of Namur (FUNDP), Namur, Belgium.

%

\end{document}